\documentclass[showpacs,twocolumn,floatfix,prb]{revtex4}
\usepackage{graphicx}
\usepackage{amsmath}
\begin{document}
\title{Degradation of electron-hole entanglement by spin-orbit coupling}
\author{J. H. Bardarson and C. W. J. Beenakker}
\affiliation{Instituut-Lorentz, Universiteit Leiden, P.O. Box 9506, 2300 RA
Leiden, The Netherlands}
\date{July 2006}
\def\bra#1{\left\langle #1 \right|}             % < |
\def\ket#1{\left| #1 \right\rangle}             % | >
\begin{abstract}
Electron-hole pairs produced by tunneling in a degenerate electron gas 
lose their spin entanglement by spin-orbit coupling, which transforms 
the fully entangled Bell state into a partially entangled mixed density 
matrix of the electron and hole spins. We calculate the dependence of 
the entanglement (quantified by the concurrence) on the spin-orbit 
coupling time $\tau_{\rm so}$ and on the diffusion time (or dwell time) 
$\tau_{\rm dwell}$ of electron and hole in the conductors (with 
conductances $\gg e^{2}/h$) at the two sides of the tunnel barrier (with 
conductance $\ll e^{2}/h$). The entanglement disappears when the ratio 
$\tau_{\rm dwell}/\tau_{\rm so}$ exceeds a critical value of order 
unity. The results depend on the type of conductor (disordered wire or 
chaotic quantum dot), but they are independent of other microscopic 
parameters (number of channels, level spacing). Our analytical treatment 
relies on an ``isotropy approximation'' (no preferential basis in spin 
space), which allows us to express the concurrence entirely in terms of 
spin correlators. We test this approximation for the case of chaotic 
dynamics with a computer simulation (using the spin kicked rotator) and 
find good agreement.
\end{abstract}
\maketitle

\section{Introduction}
Spin-orbit coupling is one of the sources of degradation of spin 
entanglement that has been extensively investigated for electron pairs 
confined to two quantum dots.\cite{Eng04} In that context the 
spin-orbit coupling induces dephasing by coupling the electron spins 
via the orbital motion to fluctuating electric fields in the environment 
(due to lattice vibrations or gate voltage fluctuations). The coupling 
of the spins to the environment is needed for entanglement degradation 
because the spin-orbit coupling by itself amounts to a local unitary 
transformation of the electron states in the two quantum dots, which cannot change the degree of entanglement.

The characteristic feature of these quantum dots is that they are
single-channel conductors with a conductance $G$ that is small compared
to the conductance quantum $e^{2}/h$. This implies in particular that
the width of the energy levels is much smaller than the mean level
spacing. At low voltages and temperatures there is then only a single
accessible orbital mode. This is the main reason that spin-orbit
coupling by itself cannot degrade the spin entanglement.

In a multi-channel conductor the situation is altogether different. Flying
qubits in a multi-channel conductor can lose their entanglement
as a result of spin-orbit coupling even in the absence of
electric field fluctuations, because the large number of
orbital degrees of freedom can play the role of an environment.
This mechanism is the electronic 
analog of the loss of polarization entanglement by 
polarization-dependent scattering in quantum optics.\cite{Vel04,Aie04,Pue06} Fully-phase-coherent 
spin-orbit coupling can degrade the spin entanglement by reducing the 
pure spin state to a mixed spin density matrix --- which typically has 
less entanglement than the pure state. Here we investigate this 
mechanism in the context of electron-hole entanglement in the Fermi sea.\cite{Bee06} Apart from the practical significance for the observability of the 
entanglement, this study provides a test for a theory of entanglement 
transfer based on the ``isotropy approximation'' that the spin state has 
no preferential quantization axis.

\begin{figure}[tb]
  \begin{center}
    \includegraphics[width=0.8\columnwidth]{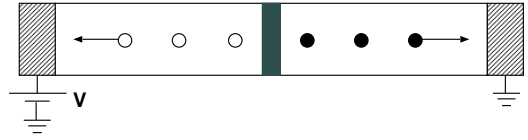} 
  \end{center}
  \caption{Multi-channel conductor containing a tunnel barrier. The applied voltage creates electron-hole pairs (solid and open
  circles) at opposite side of the barrier, whose spin state is maximally entangled. As the pair moves through the leads, the spin and
  orbital degrees of freedom become entangled by the spin-orbit coupling, degrading the spin entanglement upon tracing out the orbital
  degrees of freedom. }
  \label{fig:system}
\end{figure}
The system we consider, a multi-channel conductor containing a tunnel barrier, is schematically depicted in Fig.~\ref{fig:system}. The
applied voltage $V$ creates, at each tunnel event, a maximally entangled electron-hole pair.\cite{Bee03} Spin-orbit
coupling in the leads entangles the spin and orbital degrees of freedom. The spin state (obtained by tracing out the orbital degrees of
freedom) is degraded from a pure state to a mixed state. The degree of entanglement of the spin state decreases and can vanish for strong
spin-orbit coupling. We consider two cases. In the first case the leads are diffusive wires while in the second
case we model the leads as two chaotic cavities. Although the first case is our primary interest, we include the second case in order to test our approximate analytical calculations
against an exact numerical simulation of the spin kicked rotator.\cite{Sch89, Bar05}

The outline of the paper is as follows. In Sec.~\ref{sec:EHstate} we calculate the density matrix of the electron-hole pairs in the regime
where the tunnel conductance
$G_\text{tunnel}$ is $\ll e^2/h$. This is the regime that the electron-hole pairs form well separated current pulses, so that their entanglement can be
measured easily.\cite{Bee06} (For
$G_\text{tunnel} \gtrsim e^2/h$ different electron-hole pairs overlap in time, complicating the detection of the entanglement.) From the density matrix we seek, in Sec.~\ref{sec:Entanglement}, the degree of entanglement as measured by the
concurrence.\cite{Woo98} For our analytical treatment we approximate the density matrix by the spin-isotropic Werner state.\cite{Wer89} The absence of a
preferential basis in spin space is a natural assumption for a disordered or chaotic system, but it needs to be tested.
For that purpose we use the spin kicked rotator, which is a stroboscopic model of a chaotic cavity.\cite{Sch89,Bar05} We conclude in
Sec.~\ref{sec:conclusion}.

\section{Calculation of the electron-hole state}
\label{sec:EHstate}
\subsection{Incoming and outgoing states}
Since the scattering of both orbital and spin degrees of freedom is elastic, we may consider separately each energy $E$ in the range
($E_{\text{F}}, E_\text{F} + eV$). For ease of notation we will omit the energy arguments in what follows. We assume zero temperature, so
the incoming state is
\begin{equation}
  \label{eq:psi_in}
  \ket{\Psi_\text{in}} = \prod_{\nu=1}^{2N}a^\dagger_{L,\nu}\ket{0}.
\end{equation}
The creation operators $a^\dagger_{L,\nu}$, $\nu = 1, \ldots, 2N$ (acting on the true vacuum $\ket{0}$) occupy the $\nu$-th channel
incoming from the left. The index $\nu$ labels both the $N$ orbital and two spin degrees of freedom. The $2N$ channels incoming from the
right (creation operators $a^\dagger_{R,\nu}$) are unoccupied in the energy range ($E_{\text{F}}, E_\text{F} + eV$). We collect the
creation and annihilation operators in vectors $\mathbf{a}_L = (a_{L,1}, a_{L,2}, \ldots, a_{L,2N})$, $\mathbf{a}_R = (a_{R,1}, a_{R,2},
\ldots, a_{R,2N})$.

The annihilation operators $b_{L,\nu}$ and $b_{R,\nu}$ of the outgoing channels are related to those of the incoming channels by the
scattering matrix
\begin{equation}
  \label{eq:Smatrix}
  \begin{pmatrix} \mathbf{b}_L \\ \mathbf{b}_R \end{pmatrix} =
  S\begin{pmatrix} \mathbf{a}_L \\ \mathbf{a}_R \end{pmatrix} =
  \begin{pmatrix} r & t^\prime \\  t & r^\prime \end{pmatrix}
  \begin{pmatrix} \mathbf{a}_L \\ \mathbf{a}_R \end{pmatrix}. 
\end{equation}
The $4N\times 4N$ unitary scattering matrix $S$ is decomposed into $2N\times 2N$ transmission and reflection matrices $t, t^\prime, r$,
and
$r^\prime$. Substitution into Eq.~\eqref{eq:psi_in} gives the outgoing state
\begin{equation}
  \label{eq:psi_out}
  \ket{\Psi_\text{out}} = \prod_{\nu=1}^{2N}\left(\sum_{\nu^\prime = 1}^{2N}\left[ b^\dagger_{L,\nu'}r_{\nu'\nu} +
  b^\dagger_{R,\nu'}t_{\nu'\nu}\right]\right)\ket{0}.
\end{equation}

\subsection{Tunneling regime}
We expand the outgoing
state~\eqref{eq:psi_out} in the small parameter $\epsilon = (h/e^2)G_\text{tunnel}$, neglecting terms of order $\epsilon$ and higher.
Since $t, t'$ are $\mathcal{O}(\epsilon^{1/2})$ while $r, r'$ are $\mathcal{O}(\epsilon^{0})$, we keep only terms linear in $t$ and $t'$. The
result is 
\begin{equation}
  \label{eq:psi_outTunneling}
  \ket{\Psi_{\text{out}}} = \ket{0_\text{F}} + \sum_{\nu,\mu}(tr^\dagger)_{\nu\mu}b^\dagger_{R,\nu}b_{L,\mu}\ket{0_\text{F}} + \mathcal{O}(\epsilon),
\end{equation}
where $\ket{0_\text{F}}$ is the unperturbed Fermi sea, 
\begin{equation}
  \label{eq:FermiSea}
  \ket{0_\text{F}} = \det(r)\prod_{\nu = 1}^{2N}b^\dagger_{L,\nu}\ket{0}.
\end{equation}
Since $rr^\dagger = \openone - \mathcal{O}(\epsilon)$, we may assume that $r$ is a unitary matrix to the order in $\epsilon$ considered.
The determinant $\det(r)$ is therefore simply a phase. 
The state~\eqref{eq:psi_outTunneling} is a superposition of the unperturbed Fermi sea and a single electron-hole excitation, consisting
of an electron in channel $\nu$ at the right and a hole in channel $\mu$ at the left.

As a check, we verify that the multi-channel result~\eqref{eq:psi_outTunneling} reduces for $N=1$ to the single-channel result
\begin{align}
  \label{eq:SingleChannel}
  \ket{\Psi_{\text{out}}^{N=1}} &= \ket{0_\text{F}} +
  \frac{1}{\det(r)}\sum_{\nu,\mu}(t\sigma_yr^T\sigma_y)_{\nu\mu}b^\dagger_{R,\nu}b_{L,\mu}\ket{0_\text{F}} \notag \\&+ \mathcal{O}(\epsilon)
\end{align}
of Ref.~\onlinecite{Bee03}. We use the identity\cite{Bee04}
\begin{equation}
  \sigma_yr^T\sigma_y= \det(r)r^\dagger,
\end{equation}
which holds for any $2\times2$ unitary matrix $r$ (with $\sigma_y$ a Pauli matrix). Hence $t\sigma_yr^T\sigma_y =
\det(r)tr^\dagger+\mathcal{O}(\epsilon)$. Substitution into  the single-channel result~\eqref{eq:SingleChannel} indeed gives the multi-channel
result~\eqref{eq:psi_outTunneling} for $N=1$. 

\subsection{Spin state of the electron-hole pair}
The spin state of the electron-hole pair is obtained from $\ket{\Psi_{\text{out}}}$ by projecting out the vacuum contribution and then tracing
out the orbital degrees of freedom. This results in the $4 \times 4$ density matrix 
\begin{equation}
  \label{eq:rho}
  \rho_{\alpha\beta,\gamma\delta} = \frac{1}{w} \sum_{n,m=1}^N(tr^\dagger)_{n\alpha,m\beta}(tr^\dagger)^*_{n\gamma,m\delta},
\end{equation}
with $w = \text{tr}(t^\dagger tr^\dagger r)$. Here $n$ and $m$ label the orbital degrees of freedom and $\alpha,\beta,\gamma$, and $\delta$ label the
spin degrees of freedom.

We assume that the tunnel resistance is much larger than the resistance of the conductors at the left and right of the tunnel barrier.
The transmission eigenvalues $T_n$ (eigenvalues of $tt^\dagger$) are then determined mainly by the tunnel barrier and will depend only
weakly on the mode index $n$. We neglect this dependence entirely, so that $T_n = T$ for all $n$, the tunneling conductance being given
by $G_\text{tunnel} = (2e^2/h) NT$. 

To obtain a simpler form for the density matrix we use the polar decomposition of the scattering matrix
\begin{equation}
  S = \begin{pmatrix} r & t^\prime \\  t & r^\prime \end{pmatrix} = 
    \begin{pmatrix} u & 0 \\  0 & v \end{pmatrix}\begin{pmatrix} \sqrt{1-\mathcal{T}} & \sqrt{\mathcal{T}} \\  \sqrt{\mathcal{T}} &
      -\sqrt{1-\mathcal{T}} \end{pmatrix}
      \begin{pmatrix} u' & 0 \\  0 & v' \end{pmatrix},
\end{equation}
where $u$, $u'$, $v$, and $v'$ are unitary matrices and $\mathcal{T}=\text{diag}(T_1,T_2,\ldots, T_{2N})$. For mode independent $T_n$'s
the matrix $\mathcal{T}$ equals $T$ times the unit matrix. Hence
\begin{equation}
  tr^\dagger = \sqrt{(1-T)T}\,U
\end{equation}
is proportional to the $2N\times 2N$ unitary matrix $U = vu^\dagger$. Substitution into the expression~\eqref{eq:rho} for the density
matrix gives 
\begin{equation}
  \label{eq:rhoChannelInd}
  \rho_{\alpha\beta,\gamma\delta} = \frac{1}{2N}\sum_{n,m=1}^{N}U_{n\alpha,m\beta}U_{n\gamma,m\delta}^*.
\end{equation}

If there is no spin-orbit coupling, the matrix $U$ is diagonal in the spin indices: $U_{n\alpha,m\beta} = \tilde{U}_{nm}\delta_{\alpha\beta}$
with $\tilde{U}$ an $N \times N$ unitary matrix.
The density matrix  then represents the maximally entangled Bell state $\ket{\psi_\text{Bell}}$,
\begin{align}
  \label{eq:rhoBell}
  (\rho_\text{Bell})_{\alpha\beta,\gamma\delta} &= \frac{1}{2}\delta_{\alpha\beta}\delta_{\gamma\delta} =
  \ket{\psi_\text{Bell}}\bra{\psi_\text{Bell}}, \\
  \label{eq:Bell}
  \ket{\psi_\text{Bell}} &= \frac{1}{\sqrt{2}}\left(\ket{\uparrow}_e\ket{\uparrow}_h + \ket{\downarrow}_e\ket{\downarrow}_h\right),
\end{align}
with $\ket{\sigma}_{e,h}$ an electron ($e$) or hole ($h$) spin pointing up ($\sigma = \uparrow$) or down ($\sigma = \downarrow$).
The state~\eqref{eq:rhoBell} is a pure state ($\rho_\text{Bell}^2 = \rho_\text{Bell}$). Spin-orbit coupling will in general degrade $\rho$ to a mixed state, with less entanglement.

\section{Entanglement of the electron-hole pair}
\label{sec:Entanglement}
We quantify the degree of entanglement of the mixed electron-hole state~\eqref{eq:rho} by means of the concurrence $C$ (which is in
one-to-one correspondence with the entanglement of formation and varies from $0$ for a nonentangled state to $1$ for a
maximally entangled state). Following Wootters\cite{Woo98} the concurrence is given by
\begin{equation}
  \label{eq:Wootters}
  C = \max\left\{0,\sqrt{\lambda_1}-\sqrt{\lambda_2}-\sqrt{\lambda_3} - \sqrt{\lambda_4}\right\},
\end{equation}
where the $\lambda_i$'s are the eigenvalues, in decreasing order, of the matrix product
$\rho(\sigma_y \otimes \sigma_y)\rho^*(\sigma_y \otimes \sigma_y)$.

In the next two subsections we calculate the concurrence numerically for a chaotic cavity using Eq.~\eqref{eq:rho} and analytically with an
isotropy approximation for the density matrix.

\subsection{Numerical simulation}
We calculate the concurrence $C$ numerically for the case that the scattering at the left and at the right of the tunnel barrier is
chaotic. (The more experimentally relevant case of diffusive scattering will be considered in the next subsection.)

\begin{figure}[tb!]
  \begin{center}
    \includegraphics[width=0.5\columnwidth]{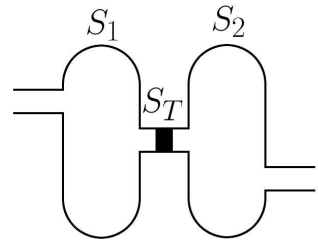} 
  \end{center}
  \caption{Two chaotic cavities with scattering matrices $S_1$ and $S_2$ connected by a tunnel barrier with scattering matrix
  $S_T$. The chaotic cavities are modeled by two spin kicked rotators.}
  \label{fig:TwoKR}
\end{figure}

The total scattering matrix $S$ of the system (shown in Fig.~\ref{fig:TwoKR}) is constructed from the scattering matrix of the tunnel barrier,
\begin{equation}
  S_T = \begin{pmatrix} \sqrt{1-T}\openone_N & \sqrt{T}\openone_N \\  \sqrt{T}\openone_N & -\sqrt{1-T}\openone_N \end{pmatrix},
\end{equation}
and the scattering matrices $S_1$ and $S_2$ of the cavity on each side of the tunnel barrier. (We denote by $\openone_N$ the $N\times
N$ unit matrix.)
We expand $S$ in the small parameter $T$ and keep terms up to order $\mathcal{O}(T^{1/2})=\mathcal{O}(\epsilon^{1/2})$,
consistent with the expansion of the outgoing state~\eqref{eq:psi_outTunneling}. This results in
\begin{subequations}
  \label{eq:rtchaotic}
\begin{align}
  r &= r_1 + t'_1\frac{1}{1-r'_1}t_1 + \mathcal{O}(T), \\
  t &= t_2\frac{1}{1+r_2}\sqrt{T}\frac{1}{1-r'_1}t_1 + \mathcal{O}(T^{3/2}),
\end{align}
\end{subequations}
and similar expressions for $r'$ and $t'$ which we do not need.

The scattering matrices $S_1$ and $S_2$ of the chaotic cavities are constructed from two spin kicked rotators. We briefly describe this
method of simulation in Appendix~\ref{app:KR}, referring to Ref.~\onlinecite{Bar05} for a more detailed exposition. 

\begin{figure}[tb!]
  \begin{center}
    \includegraphics[width=0.6\columnwidth,angle=270]{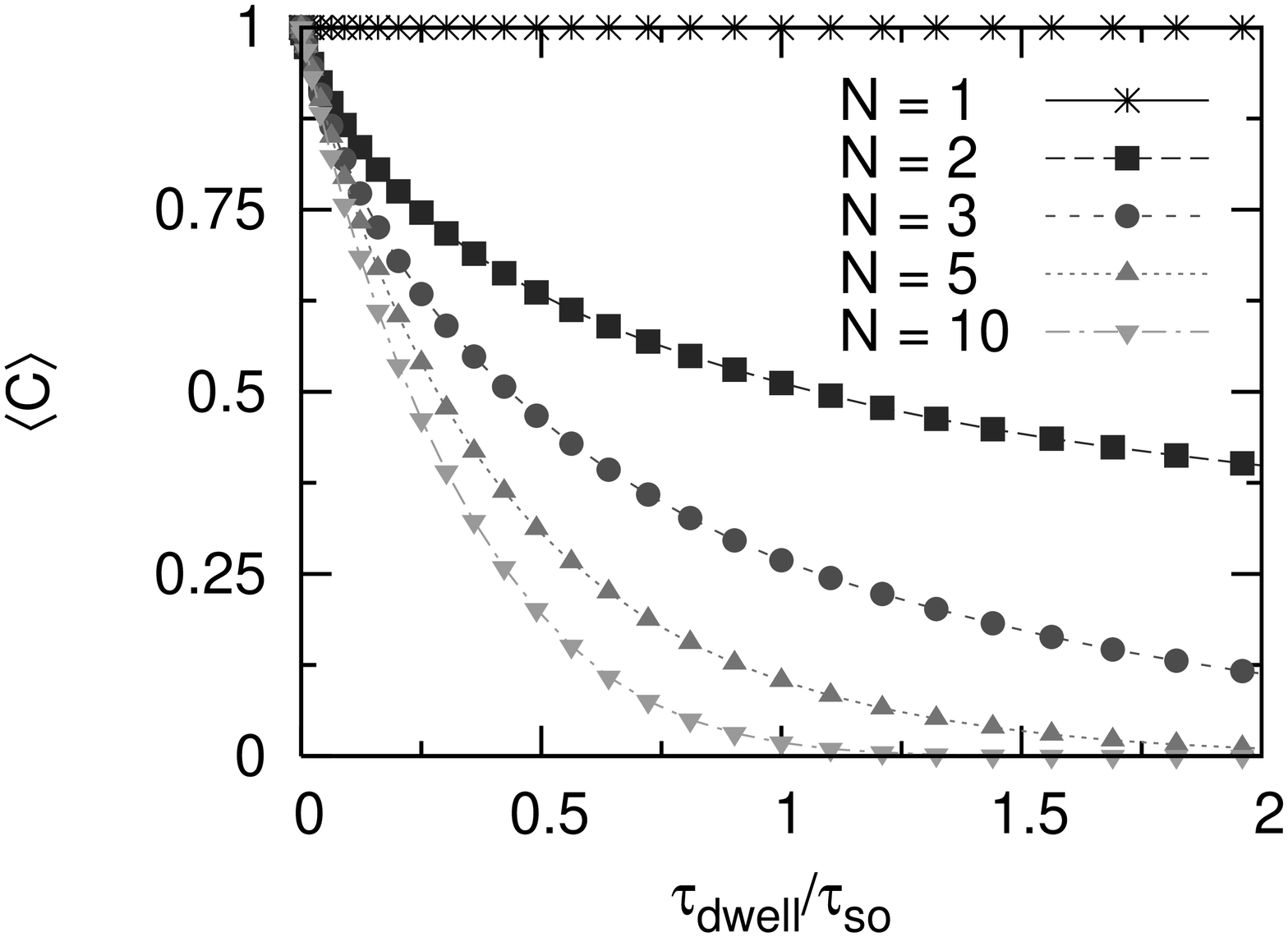} 
    \includegraphics[width=0.6\columnwidth,angle=270]{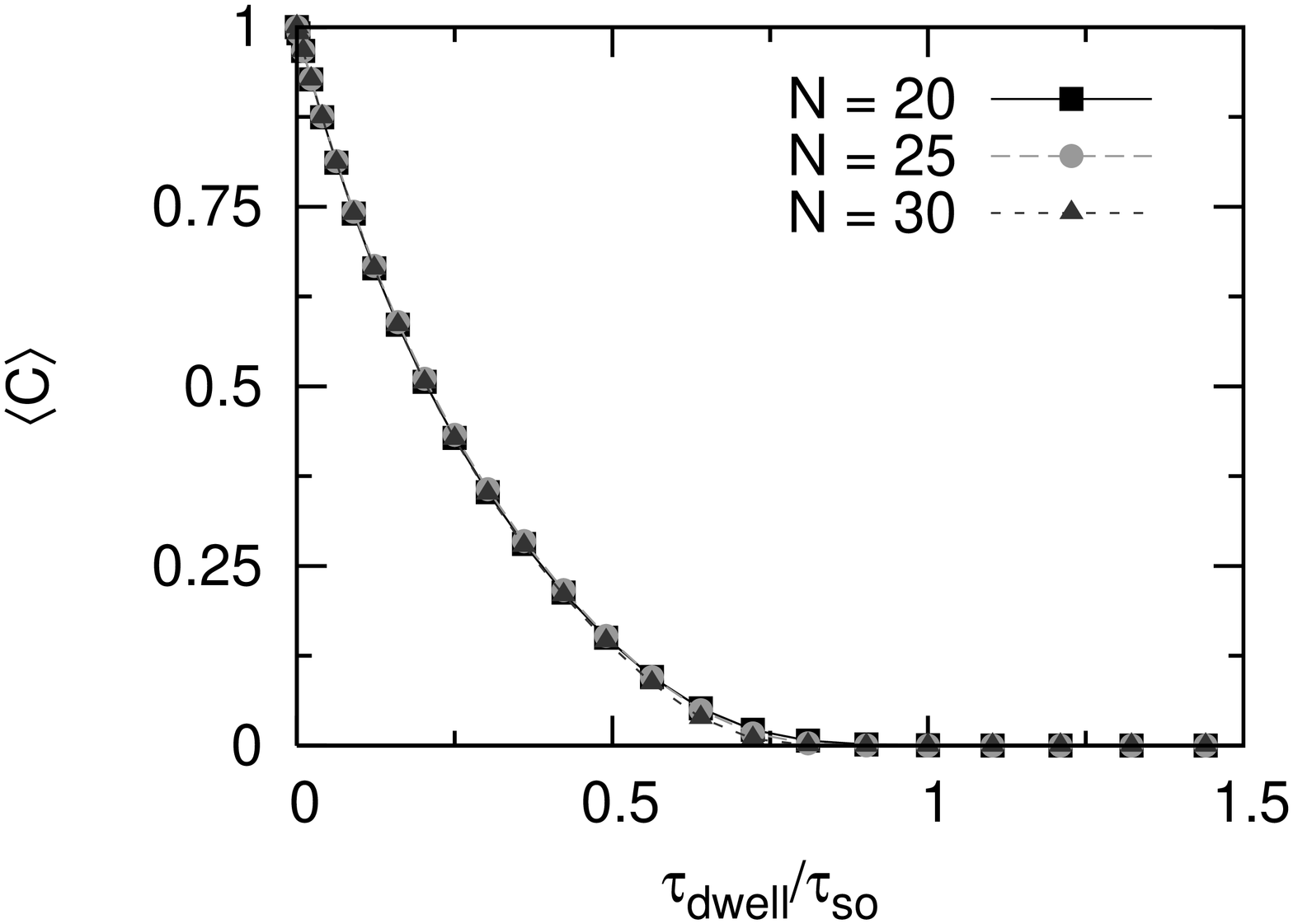}
    \includegraphics[width=0.6\columnwidth,angle=270]{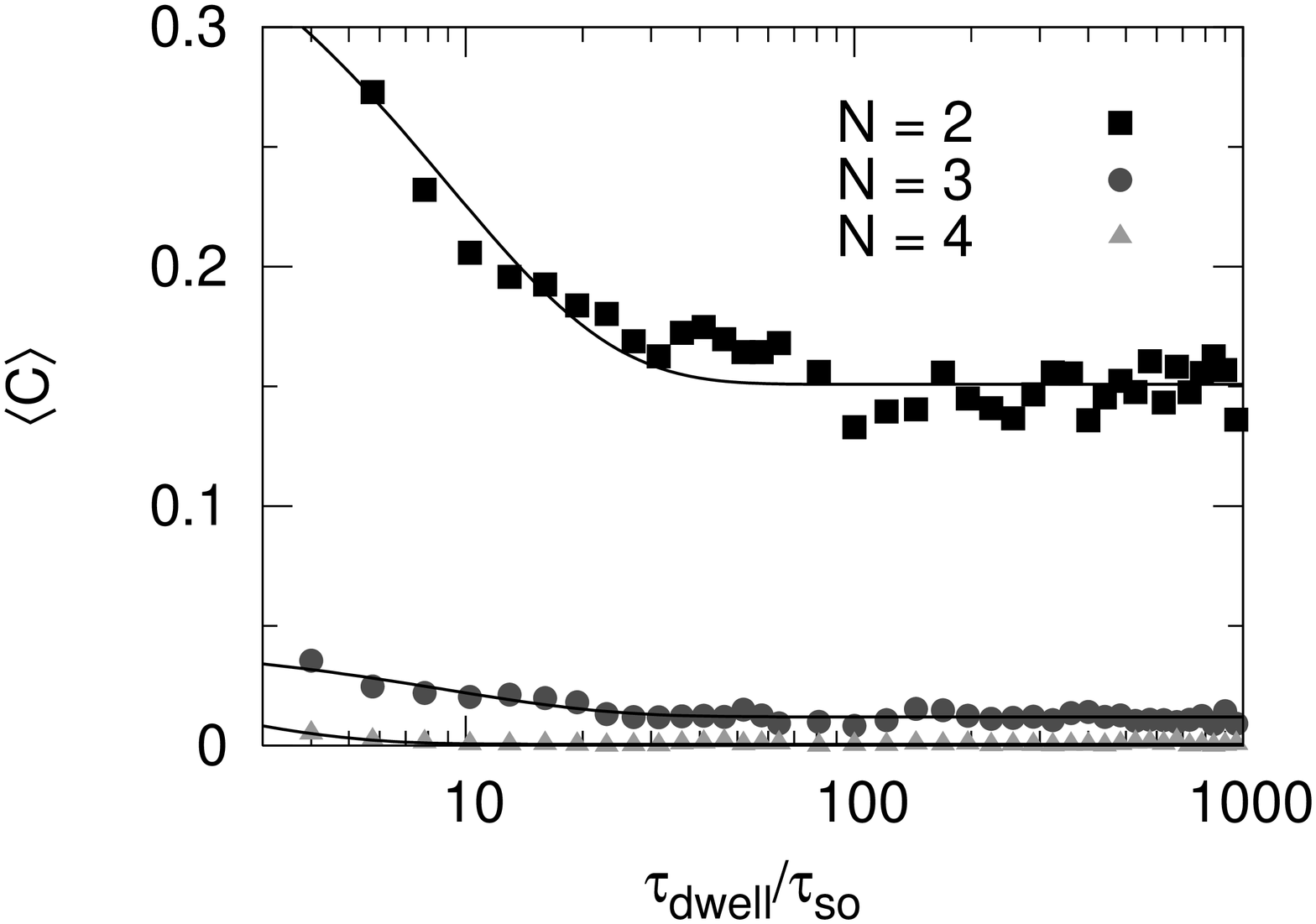}
  \end{center}
  \caption{Ensemble averaged concurrence of the electron-hole pair scattered by two chaotic cavities, as a function of the spin-orbit
  coupling rate $1/\tau_\text{so}$ for different number of modes, $N$, in the leads. The data points are calculated from the spin kicked
  rotator; the lines are guides to the eye. The middle panel shows that the results become $N$-independent for large $N$ while the bottom
  panel shows that for small $N$ the concurrence saturates at a finite value.}
  \label{fig:Cfull}
\end{figure}

The resulting ensemble-averaged concurrence as a function of the ratio $\tau_\text{dwell}/\tau_\text{so}$ of the mean dwell time $\tau_\text{dwell}$ and spin-orbit coupling time
$\tau_\text{so}$ is shown in
Fig.~\ref{fig:Cfull}. 
The dwell time $\tau_{\rm dwell}$ is the average time between a tunnel
event and the escape of the particle into the left or right reservoir.
The time $\tau_{\rm so}$ is the exponential relaxation time of the
spin-up and spin-down densities towards the equilibrium distribution.
(Both time scales are calculated in Appendixes~\ref{app:KR} and~\ref{app:analytic}.)
For a single channel, $N=1$, the concurrence is unity independent of spin-orbit coupling strength since the trace over the orbital
degrees of freedom leaves $\rho$ unchanged. From Fig.~\ref{fig:Cfull} (bottom panel) we see that for small $N$ the concurrence saturates at a nonzero value for large $\tau_\text{dwell}/\tau_\text{so}$:
\begin{equation}
  \lim_{\tau_\text{dwell}/\tau_\text{so}\rightarrow \infty} \langle C \rangle =
  \begin{cases}
    1, &N=1, \\
    0.15, &N=2, \\
    0.01, &N=3. 
  \end{cases}
\end{equation}
The limiting value for $N=2$ is close to that obtained in Ref.~\onlinecite{Fru06} in a single chaotic cavity. For $N \gtrsim 5$ the
ensemble-averaged concurrence is negligible for large $\tau_\text{dwell}/\tau_\text{so}$. The
dependence of $\langle C \rangle$ on $\tau_\text{dwell}/\tau_\text{so}$ becomes $N$ independent for $N \gtrsim 15$. 

\subsection{Isotropy approximation}
\label{sec:Werner}
To obtain an analytical expression for the entanglement degradation we approximate the density matrix by the spin-isotropic Werner
state.\cite{Wer89}
The absence of a preferential basis in spin space means that the density matrix $\rho$ for an electron-hole pair is invariant under the
transformation
\begin{equation}
  \label{eq:isotropic}
  (V \otimes V^*) \rho (V^\dagger \otimes V^T) = \rho
\end{equation}
for all $2\times 2$ unitary matrices $V$. This transformation rotates the spin basis of the electron (acted on by $V$) and the hole (acted on by $V^*$) by the
same rotation angle. The isotropy relation~\eqref{eq:isotropic} constrains the density matrix to be of the Werner form
\begin{equation}
  \label{eq:Werner_eh}
  \rho_\text{W} = \frac{1}{4}(1-\xi)\openone_4 + \xi\ket{\psi_\text{Bell}}\bra{\psi_\text{Bell}}, \quad -\frac{1}{3} \le \xi \le 1,
\end{equation}
with $\ket{\psi_\text{Bell}}$ the Bell state defined in Eq.~\eqref{eq:Bell}.
The concurrence of the electron-hole Werner state is given by
\begin{equation}
  \label{eq:CWerner}
  C(\rho_\text{W}) = \frac{3}{2}\max\left(0,\xi-\frac{1}{3}\right).
\end{equation}

The parameter $\xi$ characterizing the Werner state can be calculated from
\begin{equation}
  \label{eq:Xi}
  \xi = \text{tr } [(\sigma_z \otimes \sigma_z)\rho] = \rho_{11}-\rho_{22} - \rho_{33} +
  \rho_{44}. 
\end{equation}
Only diagonal elements of the density matrix appear in the expression~\eqref{eq:Xi} for $\xi$. These can be calculated semiclassically in the $N$-independent limit 
$N\gg 1$ (see Appendix~\ref{app:analytic}), leading to the following expressions for the concurrence: 
\begin{subequations}
  \label{eq:Canalytical}
\begin{align}
  \langle C \rangle_\text{diffusive} &=
  \begin{cases}
    \frac{3}{2}[\sum_{n=0}^\infty\xi_n]^2 - \frac{1}{2}, & 1.5\,\tau_\text{dwell} < \tau_\text{so}, \\
    0, & 0 < \tau_\text{so} < 1.5\,\tau_\text{dwell},
  \end{cases} \notag\\
  \xi_n &= \frac{4\pi(-1)^n(2n+1)}{\pi^2(2n+1)^2+8\tau_\text{dwell}/\tau_\text{so}},  \\
  \langle C \rangle_\text{chaotic} &=
  \begin{cases}
    \frac{3}{2}(1+\tau_\text{dwell}/\tau_\text{so})^{-2} - \frac{1}{2},  &\frac{\tau_\text{dwell}}{\sqrt{3}-1} < \tau_\text{so}, \\
    0, & 0 \le \tau_\text{so} \le \frac{\tau_\text{dwell}}{\sqrt{3}-1}. 
  \end{cases}
\end{align}
\end{subequations}

\begin{figure}[tb]
  \begin{center}
    \includegraphics[width=0.6\columnwidth,angle=270]{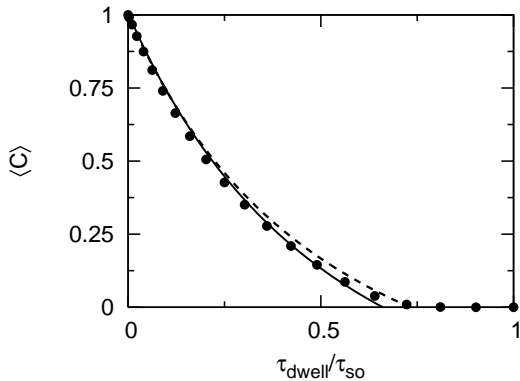} 
  \end{center}
  \caption{Ensemble averaged concurrence as a function of spin-orbit coupling rate $1/\tau_\text{so}$. The
  solid and dashed curves are the analytical
  results~\eqref{eq:Canalytical} in the case of diffusive wires and chaotic cavities, respectively, on each side of
  the tunnel barrier. These analytical curves use the isotropy approximation.
  The data points are from the numerical simulation in the chaotic case without the isotropy approximation (spin
  kicked rotator of  Fig.~\ref{fig:Cfull}, with $N=30$).}
  \label{fig:Cwerner}
\end{figure}

In Fig.~\ref{fig:Cwerner} we plot the analytical result~\eqref{eq:Canalytical} for the concurrence. The two cases of diffusive and chaotic scattering differ
only slightly. The initial slopes are the same,
\begin{align}
  \label{eq:Csmalltso}
 \langle C\rangle_\text{diffusive} &=  \langle C\rangle_\text{chaotic} \\
&= 1 - 3\tau_\text{dwell}/\tau_\text{so} + \mathcal{O}(\tau_\text{dwell}/\tau_\text{so})^2. \notag
\end{align}
The critical spin-orbit coupling strengths, beyond which the concurrence vanishes, are different:
$\tau^\text{critical}_{so} = 1.5\, \tau_\text{dwell}$ for
diffusive scattering and $\tau^\text{critical}_{so} = \tau_\text{dwell}/(\sqrt{3}-1) = 1.37\,\tau_\text{dwell}$ for chaotic scattering. 

We also compare in Fig.~\ref{fig:Cwerner} the analytical results in the chaotic case from this section with the numerical results from the previous section.
The agreement is quite good for large $N$, where the semiclassical analytics is expected to hold.

\section{Conclusion}
\label{sec:conclusion}
Figure~\ref{fig:Cwerner} summarizes our main findings: The effect of spin-orbit coupling 
on the degree of spin-entanglement of the electron-hole pairs produced 
at a tunnel barrier depends strongly on the ratio of the dwell time 
$\tau_{\rm dwell}$ and spin-orbit coupling time $\tau_{\rm so}$. Even though $\tau_{\rm dwell}$ and $\tau_{\rm so}$   
each depend sensitively on the nature of the dynamics (diffusive or 
chaotic) the dependence of the concurrence on the ratio $\tau_{\rm dwell}/\tau_{\rm so}$ is insensitive to the nature of the
dynamics. The initial decay~\eqref{eq:Csmalltso} is the same and the critical spin-orbit 
coupling strength (beyond which the entanglement vanishes) differs by 
less than 10\%. This has the useful experimental implication that a 
single parameter suffices to quantify the amount of entanglement 
degradation by spin-orbit coupling.

We have tested our analytical theory using a computer simulation for the 
case of chaotic dynamics. (The close similarity to the diffusive results 
suggests that this test is representative.) Analytics and numerics are 
in good agreement, differing by less than 10\% in the regime $N\gg 1$ of 
large conductance $G$ where the semiclassical analytics applies. While 
the semiclassical approximation is controlled by the small parameter 
$1/N$ (or, more generally, $h/e^{2}G$), the isotropy approximation has 
no small parameter that controls the error. Its use is justified by the 
reasonable expectation that an ensemble of disordered or chaotic systems 
should have no preferential quantization axis for the electron and hole 
spins. It is gratifying to see that the numerics supports this expectation.

The standard method of experimentally verifying the presence of entanglement is by demonstrating violation of Bell inequalities. In optics
this is achieved by measuring coincidence rates of photons by photodetectors (i.e.\ by counting photons) in different polarization bases.
In the solid state one cannot simply count electrons, but rather needs to formulate the Bell inequalities in terms of correlators of spin
currents (= spin noise).\cite{Kaw01, Cht02, Sam03} This has so far not been accomplished experimentally.
Thus, the isotropy approximation that has been used here as a way to simplify the 
calculation of the concurrence, also has an experimental 
implication:\cite{Mic06} By relying on spin 
isotropy the concurrence can be obtained directly from correlators of 
time averaged spin currents. Our demonstration of the accuracy of the isotropy 
approximation may motivate experimentalists to try this ``poor man's 
method'' of entanglement detection --- since average spin currents have 
been measured\cite{Jed02} while spin noise has not.

\section*{ACKNOWLEDGMENTS}
This work was supported by the Dutch Science Foundation NWO/FOM. We acknowledge support by the European Community's Marie Curie
Research
Training Network under contract No.\ MRTN-CT-2003-504574, Fundamentals of Nanoelectronics.

\appendix

\section{Spin kicked rotator}
\label{app:KR}
We use the spin kicked rotator to model transport in chaotic cavities in the presence of spin orbit coupling. The model is described in
detail in Ref.~\onlinecite{Bar05}. Here we collect the most important results of that paper for reference.

The spin kicked rotator is 
a numerically efficient stroboscopic description of the chaotic dynamics, described by a quantum map with Floquet matrix\cite{Izr90}
\begin{equation}
\mathcal{F}_{ll'}=(\Pi U X U^\dagger \Pi)_{ll'}, \quad l,l' = 0,1,\ldots, M-1.
\end{equation}
The matrices appearing in this expression are defined by
\begin{subequations}
\begin{align}
\Pi_{ll'}&=\delta_{ll'}e^{-i\pi (l+l_0)^2/M }\openone_2, \\ 
U_{ll'}&=M^{-1/2}e^{-i2\pi ll'/M}\openone_2,\\
X_{ll'}&= \delta_{ll'}e^{-i(M/4\pi)V(2\pi l/M)},
\end{align}
\end{subequations}
with kicking potential
\begin{equation}
V(\theta) =  K\cos\theta\,\openone_2 + K_\text{so}(\sigma_x\sin2\theta + \sigma_z\sin\theta).
\end{equation}
The matrices $\sigma_x$ and $\sigma_z$ are Pauli matrices. The dimension $2M \times 2M$ of
the Floquet matrix determines the mean spacing $\delta = 2\pi/M$ of the quasienergies $\varepsilon$.
The map is classically chaotic for kicking strength $K \gtrsim 7.5$. The parameter 
$K_\text{so}$ breaks spin rotation symmetry and $l_0$ breaks other symmetries of the map. In our simulations we choose $K=41$ (fully
chaotic), $M=640$, and $l_0 = 0.2$. The spin-orbit coupling time $\tau_\text{so}$ (in units of the stroboscopic period) in the model is determined by the parameter
$K_\text{so}$ through\cite{Kso}
\begin{equation}
  \tau_\text{so} = \frac{16\pi^2}{K_\text{so}^2M^2}.
\end{equation}

The reason for denoting the variable of the kicking potential by $\theta$ is historical: In its original incarnation the model was a
freely moving rotator (pendulum) periodically kicked by the kicking potential $V$ with $K_\text{so} = 0$.\cite{Izr90} The Floquet matrix was then derived from the
Hamiltonian of this kicked rotator. For our purposes it is much more fruitful to consider the Floquet matrix (time evolution
operator) $\mathcal{F}$ as the definition of the model. One then considers the matrix $X$ to describe the free (spin-orbit coupled) motion
inside the chaotic cavity. This free motion is interrupted by boundary scattering which is given in terms of the matrix $\Pi$. (The matrix
$X$ is diagonal in $\theta$ space, while $\Pi$ is diagonal in $p$-space; $U$ maps between the two spaces. $\theta$ and $p$ are conjugate
variables.) With this interpretation, the variable $\theta$ becomes the momentum-like variable, while $p$ becomes a variable for the position
on the boundary. To accommodate the notation it can be useful to think of $\theta$ as the angle describing the direction of the momentum.

\begin{figure}[tb!]
  \begin{center}
    \includegraphics[width=0.6\columnwidth]{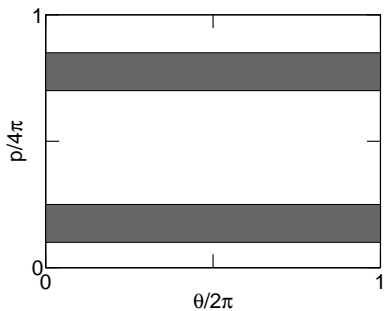} 
  \end{center}
  \caption{Location in classical phase space of the open boundary conditions representing the leads.}
  \label{fig:leads}
\end{figure}

The Floquet matrix $\mathcal{F}$ represents the internal dynamics of the chaotic cavity. In order to use the map to describe an open
cavity, we need to open up the map. This is achieved by imposing open boundary conditions
in a subspace of Hilbert space; that is, any time the map takes the particle to this subspace it is absorbed. With the above interpretation in mind, it becomes intuitively clear that these open boundary conditions take
the form of strips parallel to the $\theta$ axis in phase space (cf.\ Fig.~\ref{fig:leads}).  We thus represent the boundary conditions by the indices $l_k^{(1)}$ for the first lead and $l_k^{(2)}$ for the second lead (the subscript $k=1,2,\ldots,N$ labels the
modes). These indices label the momentum eigenstates defining the subspace. The $4N \times 2M$  projection matrix
\begin{equation}
 P_{k\alpha,k'\beta} =
  \begin{cases}
    \delta_{\alpha\beta} &\text{if } k' = l_k^{(\mu)}, \\
    0  &\text{otherwise},
  \end{cases}
\end{equation}
projects onto the leads.
The scattering matrices of the two chaotic cavities now take the form\cite{Fyo00,Oss03,Jac03,Two03}
\begin{subequations}
\begin{align}
S_1&=P[e^{-i\varepsilon}-{\mathcal F}(1-P^TP)]^{-1}{\mathcal F}P^T,\\
S_2&=P'[e^{-i\varepsilon'}-{\mathcal F}(1-P'^TP')]^{-1}{\mathcal F}P'^T,
\end{align}
\end{subequations}
where $P$ and $\varepsilon$ refer to the first chaotic cavity and $P'$ and $\varepsilon'$ refer to the second chaotic cavity. The mean dwell
time $\tau_\text{dwell}$ (in units of the stroboscopic period) is given by
\begin{equation}
  \tau_\text{dwell} = \frac{M}{N}  = \frac{2\pi}{N\delta},
\end{equation}
where we have taken into account the fact that $N$ of the $2N$ channels are closed by the tunnel barrier (cf.\ Fig.~\ref{fig:TwoKR}).
Notice that the mean dwell time is a classical quantity, while $N$ and $\delta$ separately are quantum mechanical quantities. 

From the reflection and transmission matrices~\eqref{eq:rtchaotic} the density matrix~\eqref{eq:rho} is obtained, 
 from which the concurrence~\eqref{eq:Wootters} follows. The concurrence is averaged over $20$ different quasienergies
 $\varepsilon$, $\varepsilon'$ and over $20$ different lead positions $P$, $P'$ in the two cavities (assumed to be independent scatterers). Results are shown in
 Fig.~\ref{fig:Cfull}.

\section{Calculation of spin correlators}
\label{app:analytic}
The diagonal elements of the density matrix appearing in the expression~\eqref{eq:Xi} for the Werner parameter $\xi$ represent spin
correlators,
\begin{equation}
  \rho_{11} = P_{\uparrow\uparrow},\,\, \rho_{22} = P_{\uparrow\downarrow},\,\, \rho_{33} = P_{\downarrow\uparrow},\,\, \rho_{44} = P_{\downarrow\downarrow}.
\end{equation}
Here $P_{\sigma\sigma'}$ is the probability that the outgoing electron has spin $\sigma$ and the outgoing hole has spin $\sigma'$.
To calculate these correlators, it is convenient to first consider only those electrons that exit after a time $t$ and those holes that
exit after a time $t'$. The time-resolved correlator $P_{\sigma\sigma'}(t,t')$ gives the desired $P_{\sigma\sigma'}$ after
integration over time,
\begin{equation}
  P_{\sigma\sigma'} = \int_{0}^\infty dt \int_{0}^\infty dt' P_{\sigma\sigma'}(t,t') P_\text{dwell}(t)P_\text{dwell}(t'),
\end{equation}
weighted by the dwell time distribution $P_\text{dwell}$ (we assume that the dwell times at the left and right of the tunnel barrier are
independent and identically distributed).

As initial condition we take 
\begin{equation}
  \label{eq:initial}
  P_{\sigma\sigma'}(0,0)=\frac{1}{2}\delta_{\sigma\sigma'},
\end{equation}
corresponding to the spin state immediately after the tunnel event. Spin-orbit coupling randomizes the spin with a rate
$1/\tau_\text{so}$, so that $P_{\sigma\sigma'}(t,t')$ decreases in time according to the rate equations
\begin{subequations}
  \label{eq:rate}
\begin{align}
  \frac{d}{dt}P_{\sigma\sigma'}(t,t') =& \frac{1}{2\tau_\text{so}}\sum_{\sigma''}[P_{\sigma''\sigma'}(t,t')-P_{\sigma\sigma'}(t,t')], \\
  \frac{d}{dt'}P_{\sigma\sigma'}(t,t') =& \frac{1}{2\tau_\text{so}}\sum_{\sigma''}[P_{\sigma\sigma''}(t,t')-P_{\sigma\sigma'}(t,t')].
\end{align}
\end{subequations}
The solution of the rate equations~\eqref{eq:rate} with the initial condition~\eqref{eq:initial} is
\begin{subequations}
  \label{eq:spin_corr}
\begin{align}
  P_{\uparrow\uparrow}(t,t')&=P_{\downarrow\downarrow}(t,t')= \frac{1}{4}+\frac{1}{4}e^{-(t+t')/\tau_\text{so}}, \\
  P_{\uparrow\downarrow}(t,t')&=P_{\downarrow\uparrow}(t,t')= \frac{1}{4}-\frac{1}{4}e^{-(t+t')/\tau_\text{so}}.
\end{align}
\end{subequations}

To complete the calculation we need the dwell time distribution. For a chaotic cavity this has the well known exponential form\cite{Bau90}
\begin{equation}
\label{eq:PdwellChaotic}
P_\text{dwell,chaotic} = \frac{1}{\tau_\text{dwell}}e^{-t/\tau_\text{dwell}},
\end{equation}
with 
\begin{equation}
  \tau_\text{dwell} = \frac{2\pi\hbar}{N\Delta} 
\end{equation}
inversely proportional to the mean level spacing $\Delta$ of Kramers degenerate levels in the cavity.
\begin{figure}[tb!]
  \begin{center}
    \includegraphics[width=0.6\columnwidth,angle=270]{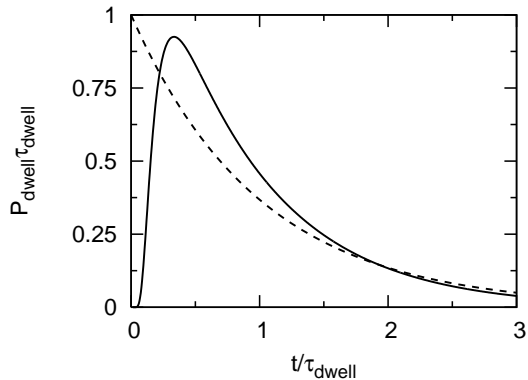} 
  \end{center}
  \caption{Dwell time distribution in a diffusive wire (solid line) and chaotic cavity (dashed line).}
  \label{fig:DwellTimeDist}
\end{figure}

For the diffusive wire (diffusion constant $D$) we determine $P_\text{dwell}$ by solving the one-dimensional
diffusion equation
\begin{equation}
    \left(\frac{\partial}{\partial t} - D\frac{\partial^2}{\partial x^2}\right)p(x,t) = 0, \quad 0<x<L,
\end{equation}
with initial and boundary conditions
\begin{equation}
    \frac{\partial p}{\partial x}(0,t) = 0, \quad p(L,t) = 0, \quad p(x,0)= \delta(x).
\end{equation}
Here $p(x,t)$ is the classical probability of finding a particle at point $x$ at time $t$. The boundary conditions represent reflection
by the high tunnel barrier at $x=0$ and absorption by the reservoir at $x=L$. 

The probability that the particle is still in the wire at time $t$ is given by
\begin{equation}
  N(t) = \int_0^L p(x,t) dx,
\end{equation}
and therefore the dwell time distribution is
\begin{equation}
  P_{\text{dwell}} = -\frac{dN(t)}{dt}.  
\end{equation}
Solution of the diffusion equation by expansion in eigenstates gives the result in the form
\begin{equation}
\label{eq:PdwellDiffusive}
  P_{\text{dwell, diffusive}}= \frac{\pi}{2\tau_\text{dwell}}\sum_{n=0}^\infty
  (-1)^n(2n+1)e^{-(2n+1)^2\frac{\pi^2}{8}\frac{t}{\tau_\text{dwell}}}. 
\end{equation}
The mean dwell time is 
\begin{equation}
  \tau_\text{dwell} = \frac{L^2}{2D}.  
\end{equation}
The dwell time distributions for the chaotic and diffusive dynamics are compared in Fig.~\ref{fig:DwellTimeDist}. 

Collecting results we arrive at the expressions~\eqref{eq:Canalytical} for the concurrence given in the main text.


\begin{thebibliography}{24}
\bibitem[{Kso()}]{Kso}
\bibinfo{note}{This expression is twice as small as in Ref.~\onlinecite{Bar05},
  because here $\tau_\text{so}$ is the relaxation time for spin densities
  whereas there it is the relaxation time for spin amplitudes.}
\end{thebibliography}
\end{document}